ORIGINAL MANUSCRIPT

# Measuring agreement among several raters classifying subjects into one or more (hierarchical) categories: A generalization of Fleiss' kappa

**Filip Moons**[1,2] · **Ellen Vandervieren**[2]

Accepted: 13 June 2025

**Abstract**

Cohen's and Fleiss' kappa are well-known measures of inter-rater agreement, but they restrict each rater to selecting only one category per subject. This limitation is consequential in contexts where subjects may belong to multiple categories, such as psychiatric diagnoses involving multiple disorders or classifying interview snippets into multiple codes of a codebook. We propose a generalized version of Fleiss' kappa, which accommodates multiple raters assigning subjects to one or more nominal categories. Our proposed $\kappa$ statistic can incorporate category weights based on their importance and account for hierarchical category structures, such as primary disorders with sub-disorders. The new $\kappa$ statistic can also manage missing data and variations in the number of raters per subject or category. We review existing methods that allow for multiple category assignments and detail the derivation of our measure, proving its equivalence to Fleiss' kappa when raters select a single category per subject. The paper discusses the assumptions, premises, and potential paradoxes of the new measure, as well as the range of possible values and guidelines for interpretation. The measure was developed to investigate the reliability of a new mathematics assessment method, of which an example is elaborated. The paper concludes with a worked-out example of psychiatrists diagnosing patients with multiple disorders. All calculations are provided as R script and an Excel sheet to facilitate access to the new $\kappa$ statistic.

## Introduction

Inter-rater agreement is the degree of agreement among independent observers who rate, code, or assess the same phenomenon. These ratings often rely on subjective evaluations provided by human raters, who sometimes differ greatly from one rater to another (Martín Andrés et al., 2025; Gwet, 2021). Various researchers in many different scientific fields have recognized this problem for a long time, since science requires measurements to be reproducible and accurate. Ideally, only a change in the subject's attribute should cause variation in the ratings, while the rater-induced source of variation should be excluded as it can jeopardize the integrity of scientific inquiries. The resolution to these problems, or at least the measurement of how big these problems are, is the study of inter-rater agreement.

The most well-known chance-corrected inter-rater agreement measures are Cohen's and Fleiss' kappa. However, these require *mutually exclusive categories*: a rater can only choose one category for each subject, and it is not possible to classify subjects into multiple categories. Remarkably, very few attempts to lift this limitation are found in the literature. In this paper, we want to develop a chance-corrected measure that allows multiple raters to classify subjects into one or more categories.

To make the kappa statistic accessible to a broad audience, this article is accompanied by a short explainer video that provides a clear, non-technical walkthrough using an Excel add-in: https://youtu.be/BxBl5qUQ8SM

✉ Filip Moons
f.moons@uu.nl

1 Freudenthal Institute, Utrecht University, PO Box 85170, 3508 AD Utrecht, the Netherlands

2 Antwerp School of Education, University of Antwerp, Antwerp, Belgium

In the rest of this introduction, we briefly introduce Cohen's kappa and Fleiss' kappa. In 'other methods', we discuss the few attempts in the literature to lift the limitation of mutually exclusive categories. In the next section, we derive the proposed measure. We start with the measure for regular categories, for which we can show that it is a generalization of Fleiss' kappa. However, the measure can easily be extended to categories that differ in importance by giving them different weights and categories that exhibit a hierarchy of interdependencies. Following this, we examine the new measure's assumptions, premises, and potential paradoxes. We also explore its range of possible values and providing guidelines for interpretation and benchmarking. Finally, we compare the proposed measure with existing methods from the literature, providing worked-out examples for illustration.

## Cohen's kappa

Starting from the 1950s, various inter-rater agreement measures have been proposed (Osgood, 1959; Bennett et al., 1954), from which Cohen's kappa (1960) is the most well-known chance-corrected measure. This correction for chance is essential, as two raters may agree by following a clear, deterministic rating procedure, or they may agree by chance (Gwet, 2012). Thus, by accounting for chance, the kappa coefficient takes into account the difficulty of the classification task at hand. The formula of Cohen's kappa is:

$$\kappa = \frac{Po - Pe}{1 - Pe}, \tag{1}$$

where $Po$ is the observed agreement and $Pe$ is the expected agreement by chance. Cohen (1960) calls the numerator the *beyond-chance*: by subtracting the observed agreement with the expected agreement by chance, you are left with 'the percent of units in which beyond-chance occurred'; the denominator $1 - Pe$ can be seen as the 'beyond-chance' in the case of perfectly agreeing raters (the observed agreement is replaced with 1). So the kappa-statistic is the proportion of the observed beyond-chance over the beyond-chance in an ideal world of perfectly agreeing raters. Hence, the $\kappa$ coefficient is the proportion of agreement *after* chance agreement is removed from consideration. $\kappa$ coefficients usually vary between $-1$ and 1, with 1 indicating perfect agreement ($Po = 1$), 0 indicating no agreement better than chance ($Pe = Po$), and a value below zero indicates the agreement was less than one would expect by chance ($Po < Pe$). The exact formulas for $Po$ and $Pe$ for the Cohen's kappa can be found in Cohen (1960).

## Fleiss' kappa

Cohen's kappa only allows to measure agreement between two independent raters, that is why Fleiss came up with the Fleiss' kappa in 1971 allowing a fixed number of two raters or more. These raters categorize subjects into exactly one of the available categories. We will now present how Fleiss defined $Po$ and $Pe$. Let $I$ be the number of subjects, $J$ is the number of raters and $C$ is the number of categories. Let $x_{ic}$ be the number of raters who classified the $i$-th subject ($i \in \{1, \dots, I\}$) into the $c$-th category ($c \in \{1, \dots C\}$). Since the categories are mutually exclusive, we know that every subject $i$ will have received exactly $J$ classifications, so $\sum_c x_{ic} = J$. We start with the observed agreement $P_o$. The extent of agreement among $J$ raters for the subject $i$ can be calculated as the proportion $P_i$ of agreeing rater pairs $\binom{x_{ic}}{2}$ out of all the $\binom{J}{2}$ possible rater pairs. If $x_{ic}$ equals 0 or 1, then there are no agreeing pairs, $\binom{x_{ic}}{2} = 0$ . This proportion $P_i$ for a subject $i$ can thus be defined as:

$$\begin{aligned} P_i &= \sum_c \frac{\binom{x_{ic}}{2}}{\binom{J}{2}} \\ &= \sum_c \frac{x_{ic}(x_{ic} - 1)}{J(J-1)} \\ &= \frac{\sum_c x_{ic}^2 - J}{J(J-1).} \end{aligned}$$

The overall observed proportion of agreement $P_o$ may then be measured by the mean of all $P_i$'s, so:

$$\begin{aligned} P_o &= \frac{1}{I} \sum_i P_i \\ &= \frac{\sum_i \sum_c x_{ic}^2 - IJ}{IJ(J-1)}. \end{aligned} \tag{2}$$

We now turn to the formula of $P_e$, the expected agreement by chance. In total, $IJ$ classifications will have been performed: all raters select exactly one category for each subject. So, the proportion of all assignments to the $c$-th category can be expressed as $\frac{\sum_i x_{ic}}{IJ}$, this is thus the probability to assign a subject to category $c$ by chance. Consequently, the probability that any pair of (independent) raters classify a subject into category $c$ by chance is given by $\left(\frac{\sum_i x_{ic}}{IJ}\right)^2$. Hence, if the raters made their classifications purely at random, the probability that two raters agree by chance on all categories is given by:

$$P_e = \sum_c \left(\frac{\sum_i x_{ic}}{IJ}\right)^2,$$

Plugging the above formulas into the $\kappa$ statistic expressed in Eq. 1, gives the Fleiss' kappa:

$$\kappa = \frac{\frac{\sum_i \sum_c x_{ic}^2 - IJ}{IJ(J-1)} - \sum_c \left( \frac{\sum_i x_{ic}}{IJ} \right)^2}{1 - \sum_c \left( \frac{\sum_i x_{ic}}{IJ} \right)^2}.$$

A more elaborate description and an example of psychiatric diagnosis on 30 subjects by six raters into a single disorder category, can be found in Fleiss (1971).

## Other methods

The literature on chance-corrected inter-rater agreement measures boomed in the 1970s and 1980s, with many proposals for different measures for different research settings. Surprisingly, only a few papers consider the limitation of mutually exclusive categories. This section briefly overviews the alternative methods in which a rater can classify a subject into multiple categories. Most of the methods below were described by Mezzich et al. (1981). R-functions implementing the methods described below are added as Supplementary material (also available at https://osf.io/q5nft/) to this article, accompanied by the worked-out examples described at the end of the article.

### Averaging or pooling Cohen's kappas

To calculate the inter-rater agreement among two raters who can classify subjects into multiple categories, a commonly used method is to calculate a Cohen's kappa for each category and average them: $\overline{\kappa}$ (De Vries et al., 2008). A problem with this approach is that when a category has an undefined Cohen's $\kappa$, $\overline{\kappa}$ is undefined too, which happens if the expected agreement by chance $Pe$ of a category is 1, e.g., when any rater did not select the category. A solution for this is *pooling* the Cohen's kappas by calculating the $Po$ and $Pe$ for each category separately and then taking the average $\overline{Po}$ and $\overline{Pe}$. Next, these averages are plugged in Eq. 1.

For example, Nvivo (2022) — a popular program for qualitative research — advocates the pooled Cohen's kappa to measure the inter-rater agreement among two coders. These two coders (= '*raters*') can code in NVivo the different sources (= '*subjects*') of their research (e.g. text fragments, interviews, pictures) to one-or-more nodes of their codebook (= '*categories*'). To get an overall $\kappa$ of this coding process, Cohen's kappa is not suited: it would only allow the coders to code a source to exactly one node in their codebook. In contrast, a source is often coded to various nodes of the codebook. Therefore, Cohen's $\kappa$ is calculated for each node in the codebook separately, and the pooled Cohen's kappa is used to get an overall $\kappa$ of the coding process (see Fig. 1).

In 2008, De Vries et al. published a simulation study in which they compared 'true' Cohen's kappa values with the (simulated) averaged kappa and the (simulated) pooled kappa. Results showed that the pooled kappa almost always deviates less from the 'true' kappa than the averaged kappa, resulting in smaller root-mean-square errors.

An important constraint to averaging or pooling Cohen's kappas is embodied in the formulas of Cohen's kappa itself: while the limitation of only one category for each subject is lifted, it is still limited to measure inter-rater agreement among exactly two raters. Moreover, it can not handle category hierarchies or different weights for categories.

### Proportional overlap

The proportional overlap method was first introduced by Mezzich et al. in 1981. The method allows the calculation of a $\kappa$ statistic in which multiple raters can classify subjects into one or more categories. The proportional overlap $\kappa$ is calculated between pairs of raters. The proportional overlap between two raters is defined as the number of categories selected by both raters divided by the total number of categories selected by either rater. For example, if a rater selected categories {blue, yellow, brown} and another rater selected {blue, green} for a given subject, their proportional overlap is the ratio of 1 (one agreement on 'blue') over 4 (in total, both raters selected four different categories: blue, yellow, brown, and green), so we get a proportional overlap of 0.25. In general, proportional overlaps range between 0 (= no overlap between the selected categories) and 1 (= perfect agreement, all categories match).

For each subject, the agreement among all raters is calculated by computing the proportional overlap for each possible pairwise combination of raters and then averaging these proportional overlaps. The overall observed agreement $P_o$ is then obtained by averaging these values for all subjects.

To determine the proportion of chance agreement $Pe$, we estimate how likely it is that two raters would select the same categories purely at random, i.e., assuming that category selections are independent of the subject being rated. Mezzich et al. (1981) approached this by considering all possible pairwise combinations of individual category selections made by raters across all subjects. For each such pair, they calculate the proportional overlap, and then average these values. This process can be implemented using software that loops over all $\binom{I \cdot J}{2}$ possible pairs, where $I$ is the number of subjects and $J$ the number of raters.

The corresponding 'Mezzich's $\kappa$' is found by plugging in $Po$ and $Pe$ in Cohen's formula (Eq. 1).

The proportional overlap method is an intuitive way to handle multiple raters classifying subjects into one or more categories and is easy to adapt to a varying number of raters (cf., some combinations of raters will not be present in this case). However, the method has limitations: it can not handle different weights for categories or category hierarchies. Moreover, the calculation of $Pe$ depends on the number of combinations $\binom{I \cdot J}{2}$, which makes computation very demanding if the number of subjects $I$ or the number of raters $J$ is high.

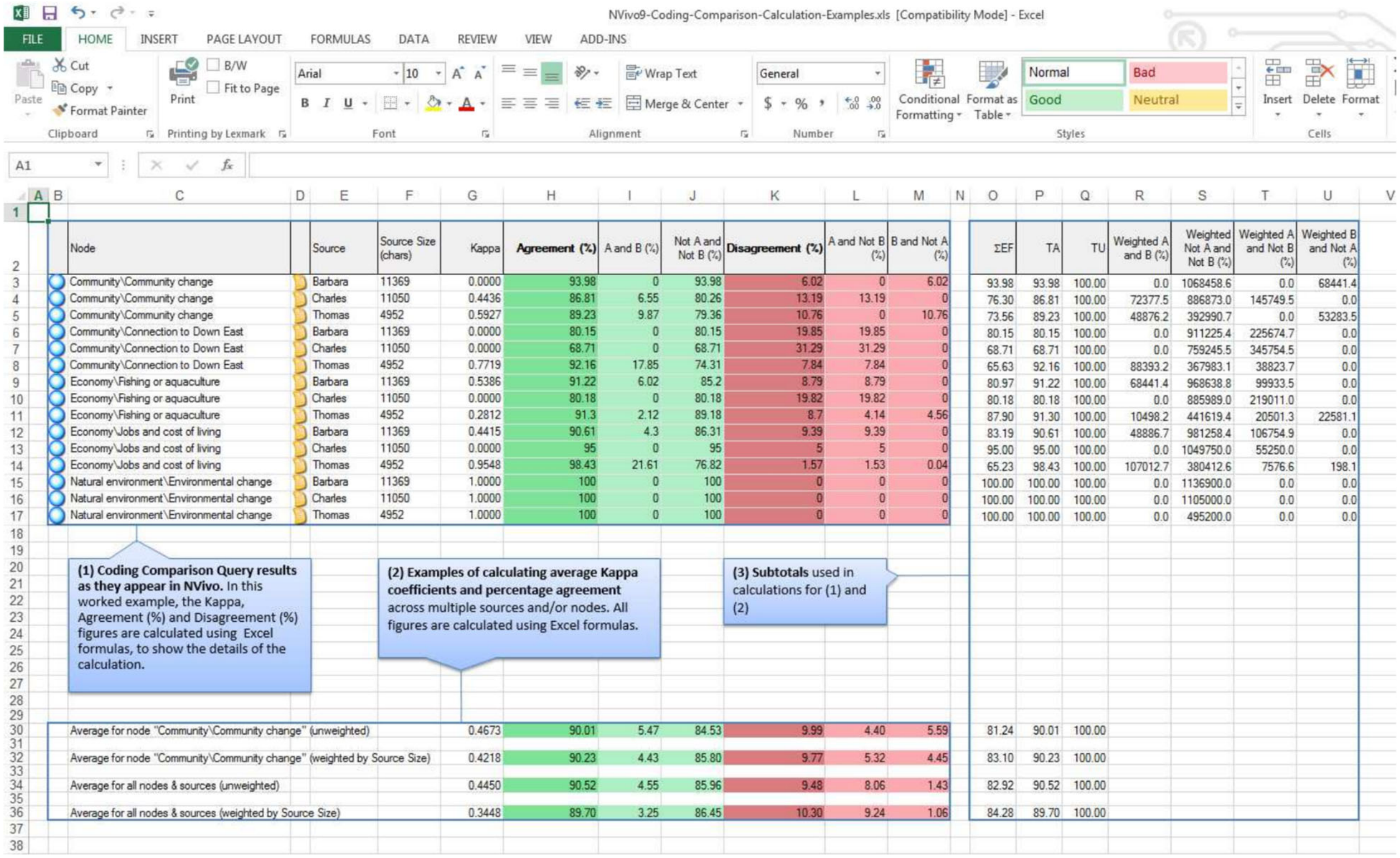

| Node | Source | Source Size (chars) | Kappa | Agreement (%) | A and B (%) | Not A and Not B (%) | Disagreement (%) | A and Not B (%) | B and Not A (%) |
|---|---|---|---|---|---|---|---|---|---|
| Community\Community change | Barbara | 11369 | 0.0000 | 93.98 | 0 | 93.98 | 6.02 | 0 | 6.02 |
| Community\Community change | Charles | 11050 | 0.4436 | 86.81 | 6.55 | 80.26 | 13.19 | 13.19 | 0 |
| Community\Community change | Thomas | 4952 | 0.5927 | 89.23 | 9.87 | 79.36 | 10.76 | 0 | 10.76 |
| Community\Connection to Down East | Barbara | 11369 | 0.0000 | 80.15 | 0 | 80.15 | 19.85 | 19.85 | 0 |
| Community\Connection to Down East | Charles | 11050 | 0.0000 | 68.71 | 0 | 68.71 | 31.29 | 31.29 | 0 |
| Community\Connection to Down East | Thomas | 4952 | 0.7719 | 92.16 | 17.85 | 74.31 | 7.84 | 7.84 | 0 |
| Economy\Fishing or aquaculture | Barbara | 11369 | 0.5386 | 91.22 | 6.02 | 85.2 | 8.79 | 8.79 | 0 |
| Economy\Fishing or aquaculture | Charles | 11050 | 0.0000 | 80.18 | 0 | 80.18 | 19.82 | 19.82 | 0 |
| Economy\Fishing or aquaculture | Thomas | 4952 | 0.2812 | 91.3 | 2.12 | 89.18 | 8.7 | 4.14 | 4.56 |
| Economy\Jobs and cost of living | Barbara | 11369 | 0.4415 | 90.61 | 4.3 | 86.31 | 9.39 | 9.39 | 0 |
| Economy\Jobs and cost of living | Charles | 11050 | 0.0000 | 95 | 0 | 95 | 5 | 5 | 0 |
| Economy\Jobs and cost of living | Thomas | 4952 | 0.9548 | 98.43 | 21.61 | 76.82 | 1.57 | 1.53 | 0.04 |
| Natural environment\Environmental change | Barbara | 11369 | 1.0000 | 100 | 0 | 100 | 0 | 0 | 0 |
| Natural environment\Environmental change | Charles | 11050 | 1.0000 | 100 | 0 | 100 | 0 | 0 | 0 |
| Natural environment\Environmental change | Thomas | 4952 | 1.0000 | 100 | 0 | 100 | 0 | 0 | 0 |
| Average for node "Community\Community change" (unweighted) | | | 0.4673 | 90.01 | 5.47 | 84.53 | 9.99 | 4.40 | 5.59 |
| Average for node "Community\Community change" (weighted by Source Size) | | | 0.4218 | 90.23 | 4.43 | 85.80 | 9.77 | 5.32 | 4.45 |
| Average for all nodes & sources (unweighted) | | | 0.4450 | 90.52 | 4.55 | 85.96 | 9.48 | 8.06 | 1.43 |
| Average for all nodes & sources (weighted by Source Size) | | | 0.3448 | 89.70 | 3.25 | 86.45 | 10.30 | 9.24 | 1.06 |

| Node | Source | ΣEF | TA | TU | Weighted A and B (%) | Weighted Not A and Not B (%) | Weighted A and Not B (%) | Weighted B and Not A (%) |
|---|---|---|---|---|---|---|---|---|
| Community\Community change | Barbara | 93.98 | 93.98 | 100.00 | 0.0 | 1068458.6 | 0.0 | 68441.4 |
| Community\Community change | Charles | 76.30 | 86.81 | 100.00 | 72377.5 | 886873.0 | 145749.5 | 0.0 |
| Community\Community change | Thomas | 73.56 | 89.23 | 100.00 | 48876.2 | 392990.7 | 0.0 | 53283.5 |
| Community\Connection to Down East | Barbara | 80.15 | 80.15 | 100.00 | 0.0 | 911225.4 | 225674.7 | 0.0 |
| Community\Connection to Down East | Charles | 68.71 | 68.71 | 100.00 | 0.0 | 759245.5 | 345754.5 | 0.0 |
| Community\Connection to Down East | Thomas | 65.63 | 92.16 | 100.00 | 88393.2 | 367983.1 | 38823.7 | 0.0 |
| Economy\Fishing or aquaculture | Barbara | 80.97 | 91.22 | 100.00 | 68441.4 | 968638.8 | 99933.5 | 0.0 |
| Economy\Fishing or aquaculture | Charles | 80.18 | 80.18 | 100.00 | 0.0 | 885989.0 | 219011.0 | 0.0 |
| Economy\Fishing or aquaculture | Thomas | 87.90 | 91.30 | 100.00 | 10498.2 | 441619.4 | 20501.3 | 22581.1 |
| Economy\Jobs and cost of living | Barbara | 83.19 | 90.61 | 100.00 | 48886.7 | 981258.4 | 106754.9 | 0.0 |
| Economy\Jobs and cost of living | Charles | 95.00 | 95.00 | 100.00 | 0.0 | 1049750.0 | 55250.0 | 0.0 |
| Economy\Jobs and cost of living | Thomas | 65.23 | 98.43 | 100.00 | 107012.7 | 380412.6 | 7576.6 | 198.1 |
| Natural environment\Environmental change | Barbara | 100.00 | 100.00 | 100.00 | 0.0 | 1136900.0 | 0.0 | 0.0 |
| Natural environment\Environmental change | Charles | 100.00 | 100.00 | 100.00 | 0.0 | 1105000.0 | 0.0 | 0.0 |
| Natural environment\Environmental change | Thomas | 100.00 | 100.00 | 100.00 | 0.0 | 495200.0 | 0.0 | 0.0 |
| Average for node "Community\Community change" (unweighted) | | 81.24 | 90.01 | 100.00 | | | | |
| Average for node "Community\Community change" (weighted by Source Size) | | 83.10 | 90.23 | 100.00 | | | | |
| Average for all nodes & sources (unweighted) | | 82.92 | 90.52 | 100.00 | | | | |
| Average for all nodes & sources (weighted by Source Size) | | 84.28 | 89.70 | 100.00 | | | | |

**Fig. 1** NVivo advocates the pooled Cohen's kappa approach in the provided Excel sheets to get an overall $\kappa$ of the coding process (Nvivo, 2022)

## Chance-corrected intraclass correlations

Mezzich et al. (1981) also proposed a method using intraclass correlation coefficients as an intermediate step in calculating a kappa statistic that allows multiple raters to classify each subject into multiple categories. To compute the intraclass correlations, let $\mathbf{x_{ij}} = (x_{ij1}, x_{ij2}, \ldots, x_{ijC})$ represent the classification vector of the $i$-th subject ($i \in \{1, \ldots, I\}$) by the $j$-th rater ($j \in \{1, \ldots, J\}$), where $x_{ijc} = 1$ if subject $i$ was classified by rater $j$ into category $c$ ($c \in \{1, \ldots, C\}$), and $x_{ijc} = 0$ otherwise.

A measure of agreement is obtained by computing an intraclass correlation coefficient $\rho_i$ across all $\mathbf{x_{ij}}$ vectors for a given subject $i$ using a one-way ANOVA.

The observed agreement $P_o$ is then calculated as the average of $\rho_1, \rho_2, \ldots, \rho_I$. The expected agreement $P_e$ is computed by calculating the intraclass correlation coefficient across all classification vectors $\mathbf{x_{ij}}$ for all raters and subjects. Plugging $P_o$ and $P_e$ into Eq. 1 yields the chance-corrected intraclass correlation coefficient. Although the method is appealing in its simplicity, it cannot accommodate different weights for categories or hierarchical category structures.

## Chance-corrected rank correlations

The method proposed by Kraemer in 1980 is the only one found in the literature where multiple raters classify subjects using an ordered list of categories: for example, the best-fitting category according to the rater is ranked first, the second-best category second, and so on.

To calculate the corresponding kappa statistic, Kraemer uses classification vectors $\mathbf{x_{ij}} = (x_{ij1}, \ldots, x_{ijC})$ that contain the ranks $x_{ijc}$ of category $c$ assigned by rater $j$ to subject $i$. In this vector, the most preferred category receives a rank of 1, the second a rank of 2, and so on. Tied ranks can also be included in $\mathbf{x_{ij}}$. Categories not selected by the rater are assigned the average of the remaining ranks.

Consider, for example, the case where a rater ranks the following categories for a subject: 1. green, 2. brown, 2. orange, 2. red, 3. yellow, based on eight available categories {blue, brown, green, pink, purple, orange, red, yellow}. Green receives rank 1, and brown, orange, and red receive rank 3 (the average of ranks 2, 3, and 4). Yellow receives rank 5.

The unselected categories – blue, pink, and purple – receive a rank of 7 (the average of the remaining ranks 6, 7, and 8). The resulting vector $\mathbf{x_{ij}}$ is (7, 3, 1, 7, 7, 3, 3, 5).

The chance-corrected rank correlation $\kappa$ is calculated between pairs of raters using the Spearman correlation coefficient, which measures agreement between two ranked classification vectors.

The calculation of $P_o$ and $P_e$ follows the same logic as in the proportional overlap method (1.3.2). For $P_o$, we first compute, for each subject, the average of all Spearman correlation coefficients across all possible pairwise combinations of raters. These values are then averaged for all subjects. For $P_e$, we compute the Spearman correlation coefficient for each pairwise combination of classification vectors made by raters across all subjects and take the average of these values. The corresponding $\kappa$ is obtained by plugging $P_o$ and $P_e$ into Eq. 1.

While this method is the only chance-corrected inter-rater agreement measure in the literature that allows for ranked classifications, it cannot accommodate different weights for categories or hierarchical structures–though these limitations may be less relevant in ranked classification contexts. The computational complexity of calculating $P_e$ is equivalent to that of the proportional overlap method.

## Derivation of the proposed kappa statistic

### Non-hierarchical categories

Suppose a sample of $I$ subjects has been classified by the same set of $J$ raters into $C$ categories. The $C$ categories are not mutually exclusive: a subject can be classified by a rater into multiple categories. Let $\mathbf{x_{ij}} = (x_{ij1}, x_{ij2}, \ldots, x_{ijC})$ represent the classification vector of the $i$-th subject ($i \in \{1, \ldots, I\}$) for the $j$-th rater ($j \in \{1, \ldots, J\}$), with $x_{ijc} = 1$ when subject $i$ was classified by rater $j$ into category $c$ ($c \in \{1, \ldots C\}$), and $x_{ijc} = 0$ otherwise. Let $x_{ic} = \sum_j x_{ijc}$ denote the number of raters classifying subject $i$ into category $c$, with $J - x_{ic}$ representing the number of raters that did not classify subject $i$ into category $c$. We can assemble all $x_{ic}$'s in an $I \times C$-matrix $\mathbf{X}$, containing all classifications. Some scholars would call $\mathbf{X}$ the 'agreement table.'

In case the categories are non-hierarchical, the selection of a category is independent from the (non-)selection of the other categories. The idea behind our proposed $\kappa$ statistic is that we first derive a kappa statistic like the one described by Cohen (1960) for each category $c$:

$$\kappa_c = \frac{Po_c - Pe_c}{1 - Pe_c}, \tag{3}$$

where $Po_c$ is the observed agreement for category $c$ and $Pe_c$ is the proportion of agreement expected by chance for category $c$. In our proposed $\kappa$ statistic, the $k_c$'s are not used directly, they solely give an impression on the agreement within each category separately. Instead, we will use the $Po_c$'s and $Pe_c$'s and pool them together into one $\kappa$ statistic.

We will calculate $Po_c$ pairwise (Conger, 1980). Two raters $a$ and $b$ agree on subject $i$ when they both classified subject $i$ into category $c$ (so $x_{iac} = x_{ibc} = 1$) or when they both did not classify subject $i$ into category $c$ (so $x_{iac} = x_{ibc} = 0$). Hence, the extent of agreement for subject $i$ and category $c$, can be seen as the proportion of rater pairs with agreement for category $c$ to the total number of rater pairs. So, for subject $i$ and category $c$, the numerator consists of the sum of $\binom{x_{ic}}{2}$ and $\binom{J-x_{ic}}{2}$, while the denominator is the number of all possible rater pairs $\binom{J}{2}$.

So the extent of agreement for subject $i$ and category $c$ can be expressed as:

$$\frac{\binom{x_{ic}}{2} + \binom{J-x_{ic}}{2}}{\binom{J}{2}}.$$

To measure the overall observed proportion of agreement $Po_c$ for category $c$, we aggregate over all subjects. The numerator sums the number of rater pairs in agreement for each subject, and the denominator sums the total number of possible rater pairs for each subject[1]:

$$\begin{aligned} Po_c &= \frac{\sum_i \left[\binom{x_{ic}}{2} + \binom{J-x_{ic}}{2}\right]}{\sum_i \binom{J}{2}} \qquad (4) \\ &= \frac{\sum_i \left[x_{ic}(x_{ic}-1) + (J - x_{ic})(J - x_{ic} - 1)\right]}{\sum_i J(J-1)} \\ &= \frac{\sum_i \left(2x_{ic}^2 - 2Jx_{ic} + J^2 - J\right)}{IJ(J-1)} \qquad (5) \end{aligned}$$

$Pe_c$ denotes the probability that two raters agree on (not) selecting category $c$ by chance. For each category $c$, $IJ$ decisions of (not) selecting $c$ will have been performed. As $x_{ic}$ denotes the number of raters classifying subject $i$ into category $c$, $\sum_i x_{ic}$ represents the total number of classifications into category $c$. Hence, the proportion $\frac{\sum_i x_{ic}}{IJ}$ equals the probability that a rater randomly classifies a subject into category $c$. In case of two (independent) raters, the probability that both raters classify a subject into category $c$ by chance is thus $\left(\frac{\sum_i x_{ic}}{IJ}\right)^2$. If $x_{ic}$ raters classified subject $i$ into category $c$, $J - x_{ic}$ raters did not. As such, the proportion $\frac{\sum_i (J-x_{ic})}{IJ}$ represents the probability that a rater did not classify a subject into category $c$ by chance. In case of two (independent) raters, the probability that both raters did not classify a subject into category $c$ by chance is thus $\left(\frac{\sum_i (J - x_{ic})}{IJ}\right)^2$. Hence, the probability that two raters agree on (not) selecting category $c$ by chance equals:

$$\begin{aligned} Pe_c &= \left(\frac{\sum_i x_{ic}}{IJ}\right)^2 + \left(\frac{\sum_i (J - x_{ic})}{IJ}\right)^2 \\ &= \left(\frac{\sum_i x_{ic}}{IJ}\right)^2 + \left(\frac{IJ - \sum_i x_{ic}}{IJ}\right)^2 \\ &= \left(\frac{\sum_i x_{ic}}{IJ}\right)^2 + \frac{I^2J^2 - 2 \cdot IJ \cdot \sum_i x_{ic} + (\sum_i x_{ic})^2}{I^2J^2} \\ &= 2\left(\frac{\sum_i x_{ic}}{IJ}\right)^2 - 2\left(\frac{\sum_i x_{ic}}{IJ}\right) + 1 \end{aligned} \quad (6)$$

[1] Note that $Po_c$ can also be derived as the average of subject-wise agreement proportions, similar to Fleiss' kappa (see Eq. 2).

We now aggregate all $Po_1, \ldots, Po_C$ and $Pe_1, \ldots, Pe_C$ into one kappa-statistic, including each category[2]:

$$\kappa = \frac{\sum_c (Po_c - Pe_c)}{\sum_c (1 - Pe_c)}. \quad (7)$$

## Adding weights to categories

It is easy to extend Eq. 7 to account for differences in the relative importance of categories. To do so, we introduce a weight vector $\mathbf{w} = (w_1, \ldots, w_C)$, where each $w_c$ reflects the importance assigned to category $c$, relative to the other categories. The specific choice of $\mathbf{w}$ should reflect the priorities or stakes relevant to the research context in which the classifications were made.

While it is often convenient to normalize the weights such that $\sum_{c=1}^{C} w_c = 1$, this is not strictly necessary for the method to work. When all categories are considered equally important – sometimes referred to as the unweighted or equally weighted case – each weight can simply be set to $w_c = 1$ for all $c$.

To incorporate these differences in importance, we multiply each category's contribution to the numerator and denominator of Eq. 7 by its corresponding weight $w_c$, yielding:

$$\kappa = \frac{\sum_c w_c(Po_c - Pe_c)}{\sum_c w_c(1 - Pe_c)}. \quad (8)$$

## The proposed $\kappa$ statistic is a generalization of Fleiss' kappa

When the requirements of the Fleiss' kappa are fulfilled, our proposed $\kappa$-static reduces to it:

**Theorem 1** *In case of equally weighted, mutually exclusive and non-hierarchical categories, the proposed kappa-statistic in Eq. 7 reduces to the Fleiss' kappa.*

A detailed proof is provided in the Supplementary material.

## Handling missing data or a varying number of raters

Until now, we only considered the case of a fixed number of raters $J$. However, in practice, raters may only have classified a proportion of the participating subjects or used only a subset of the available categories. Two possibilities can be distinguished:

1. **Missing data**: some classifications are lost due to unforeseen circumstances. However, the experiment was originally designed to collect this data.
2. **Varying number of raters**: raters were only expected to rate a subset of subjects or categories, due to practical or feasibility constraints.

To account for these cases, we replace the fixed number of raters $J$ with an $I \times C$ matrix $\mathbf{J}'$, whose elements $j_{ic}$ represent the number of raters who had the opportunity to classify subject $i$ into category $c$. The observed agreement $Po_c$ from Eq. 4 then becomes:

$$\begin{aligned} Po_c &= \frac{\sum_i \left[\binom{x_{ic}}{2} + \binom{j_{ic} - x_{ic}}{2}\right]}{\sum_i \binom{j_{ic}}{2}} \\ &= \frac{\sum_i (2x_{ic}^2 - 2j_{ic}x_{ic} + j_{ic}^2 - j_{ic})}{\sum_i j_{ic}(j_{ic} - 1)}, \end{aligned}$$

and for $Pe_c$ (Eq. 6), we get:

$$\begin{aligned} Pe_c &= \left(\frac{\sum_i x_{ic}}{\sum_i j_{ic}}\right)^2 + \left(\frac{\sum_i (j_{ic} - x_{ic})}{\sum_i j_{ic}}\right)^2 \\ &= 2\left(\frac{\sum_i x_{ic}}{\sum_i j_{ic}}\right)^2 - 2\left(\frac{\sum_i x_{ic}}{\sum_i j_{ic}}\right) + 1. \end{aligned}$$

[2] Note that the proposed kappa statistic in Eq. 7 is not an average of the individual $k_c$'s (Eq. 3). Simulations indicate that pooling the $Po_c$'s and $Pe_c$'s in this way results in smaller root-mean-square errors compared to using a weighted average of the $k_c$'s. These simulations are similar to the study by De Vries et al. (2008) that compared averaging or pooling Cohen's kappas (see 1.3.1). In addition to the smaller root-mean-square errors, this aggregation mechanism makes the $\kappa$ statistic insensitive to undefined $\kappa_c$ values (e.g., when any rater did not select it; see Section "Insensitivity to unused categories and the 'always-selected category'-paradox") and makes it a true generalization of Fleiss' kappa (see Section "The proposed $\kappa$ statistic is a generalization of Fleiss' kappa").

Although the proposed $\kappa$ statistic is flexible enough to handle missing classifications in some categories with the formulas above, this situation is often undesirable from a methodological perspective. If raters do not have access to the full set of categories, their classifications may differ from those they would have made under complete information. Therefore, it is generally preferable to apply these adjustments only when the number of raters varies across subjects, not categories. In that case, the matrix $\mathbf{J}'$ can be replaced by a vector $\mathbf{j} = (j_1, j_2, \ldots, j_I)^T$, with $j_i$ the number of raters who classified subject $i$. The formulas for $Po_c$ (Eq. 5) and $Pe_c$ (Eq. 6) then simplify to:

$$Po_c = \frac{\sum_i (2x_{ic}^2 - 2j_i x_{ic} + j_i^2 - j_i)}{\sum_i j_i(j_i - 1)}, \tag{9}$$

$$Pe_c = 2\left(\frac{\sum_i x_{ic}}{\sum_i j_i}\right)^2 - 2\left(\frac{\sum_i x_{ic}}{\sum_i j_i}\right) + 1 \tag{10}$$

## Hierarchical categories

### Actual classifications versus possible classifications

Let us now consider the case when categories have some kind of hierarchical structure. For example, the categories to which a rater classifies subjects can have main categories and subcategories, with a subcategory only being selectable if the main category was chosen. More complex hierarchical structures are also possible: think of decision graphs in which some subcategories can only be chosen when certain conditions are met (e.g., a category can be selected only when one of two other categories is selected, or only when another is not selected).

No matter how the hierarchical structure of the categories is constructed, all these hierarchies have one thing in common: based on the classifications rater $j$ already made for subject $i$, some (sub)categories will (not) be selectable. In other words, whereas in the non-hierarchical case every subject $i$ could be classified $J$ times into category $c$, in the hierarchical case the upper limit of possible classifications will depend on the number of raters who could select category $c$, we will denote these *possible classifications* as $s_{ic}$. This ensures that the agreement statistic only considers the number of raters for whom category $c$ was actually available, avoiding unfair penalization for categories that were not selectable due to the hierarchy.

It is important to understand the difference between the $s_{ic}$ and $x_{ic}$ values for a given subject $i$ and category $c$: by definition, $x_{ic} \leq s_{ic}$ and $s_{ic} \in \{0, 1, \ldots, J\}$. $x_{ic}$ denotes the number of *actual* classifications of subject $i$ into category $c$; so the number of times category $c$ was selected for subject $i$, while $s_{ic}$ indicates the number of *possible* classifications of subject $i$ into category $c$. This means that $s_{ic}$ corresponds to the number of times category $c$ was available for selection for subject $i$, which directly follows from the hierarchical structure of the categories. The calculation of $s_{ic}$ for a given category $c$ and subject $i$ can depend on actual classifications of higher-order categories for subject $i$, but never on $x_{ic}$ itself. For convenience, we can collect all $s_{ic}$ elements in an $I \times C$-matrix $\mathbf{S}$. We will see that taking the hierarchy of the categories into account depends solely on these $s_{ic}$ values in the computation of the $\kappa$ statistic.

To give an impression of how to calculate the $s_{ic}$'s in a simple parent–child hierarchical structure: all main categories could be selected by all $J$ raters for every subject $i$, so $s_{ic} = J$ for all main categories. A child category $c'$ can only be selected if the parent category $p$ was selected, so $s_{ic'} = x_{ip}$, i.e., the number of *possible* classifications into child category $c'$ for subject $i$ equals the number of *actual* classifications into parent category $p$. For more complex hierarchical structures, the calculation of $s_{ic}$ can depend on several different $x_{ijc}$ values and may involve the inclusion–exclusion principles of combinatorics.

A worked-out example illustrating the calculation of $s_{ic}$ for both simple and hierarchical category structures is provided as a first example in Section "Worked-out examples".

### The kappa-statistic

With the introduction of the matrix $\mathbf{S}$, the construction of $Po_c$ and $Pe_c$ becomes straightforward: we replace each occurrence of $J$ in Eqs. 4 and 6 with the corresponding $s_{ic}$ values. This yields:

$$\begin{aligned} Po_c &= \frac{\sum_i \left[\binom{x_{ic}}{2} + \binom{s_{ic}-x_{ic}}{2}\right]}{\sum_i \binom{s_{ic}}{2}} \\ &= \frac{\sum_i [(x_{ic})(x_{ic}-1) + (s_{ic}-x_{ic})(s_{ic}-x_{ic}-1)]}{\sum_i s_{ic}(s_{ic}-1)} \\ &= \frac{\sum_i (2x_{ic}^2 - 2s_{ic}x_{ic} + s_{ic}^2 - s_{ic})}{\sum_i s_{ic}(s_{ic}-1)}, \end{aligned} \tag{11}$$

and for $Pe_c$:

$$\begin{aligned} Pe_c &= \left(\frac{\sum_i x_{ic}}{\sum_i s_{ic}}\right)^2 + \left(\frac{\sum_i s_{ic} - x_{ic}}{\sum_i s_{ic}}\right)^2 \\ &= 2\left(\frac{\sum_i x_{ic}}{\sum_i s_{ic}}\right)^2 - 2\left(\frac{\sum_i x_{ic}}{\sum_i s_{ic}}\right) + 1. \end{aligned} \tag{12}$$

If we were to aggregate $Po_c$ and $Pe_c$ in the same way as in Eq. 8, we would have adjusted the contribution of category $c$ according to the context-related weights $w_c$. However, in this aggregation, the contribution of category $c$ would not be adjusted relative to its total number of possible classifications, $\sum_i s_{ic}$, in the overall $\kappa$ calculation. Although $Po_c$ and $Pe_c$ are correctly calculated based on the possible classifications $s_{ic}$, this does not ensure that the overall contribution of category $c$ to $\kappa$ is proportionate to its hierarchical availability across subjects and raters.

This is not desirable, as the following example illustrates: consider unweighted categories, and assume that for a subject $i$ only two raters could select subcategory $c'$, so $s_{ic'} = 2$. Rater 1 classified subject $i$ into subcategory $c'$, and rater 2 did not. Moreover, due to the category hierarchy, the subcategory $c'$ was not available for any other subjects across all raters, so $\sum_i s_{ic'} = 2$. This will lead to a $Po_{c'} = 0$ and $Pe_{c'} = 0.5$. Without further scaling for the total possible occurrences of a category (and thus using formula Eq. 8 for aggregating $Po_c$ and $Pe_c$), the subcategory will contribute $-0.5$ to the numerator and 0.5 to the denominator of $\kappa$. In other words, if we do not adjust the contribution of $c'$ for the number of possible classifications, we risk pulling the value of $\kappa$ down due to an almost negligible category that was only selectable on two occasions. In contrast, the main categories had $IJ$ possible classifications.

To solve the problem and adjust for the total possible classifications $s_{ic}$ of category $c$, we introduce a scaling factor $\phi_c$ for each category $c$, to scale the terms $Po_c - Pe_c$ in the numerator and the terms $1 - Pe_c$ in the denominator:

$$\phi_c = \frac{\sum_i s_{ic}}{IJ}. \tag{13}$$

This scaling factor contrasts the total possible occurrences of a category with the $IJ$ possible classifications of main categories. As a result, the main categories always have $\phi_c = 1$. With the expressions in Eqs. 11, 12 and 13, we are now ready to define the kappa-statistic for the hierarchical case:

$$\kappa = \frac{\sum_c w_c \phi_c (Po_c - Pe_c)}{\sum_c w_c \phi_c (1 - Pe_c)}. \tag{14}$$

### Handling missing data or a varying number of raters

Note that in the calculation of the proposed kappa-statistic for hierarchical categories Eq. 14, only the scaling factors $\phi_c$ still refer to the assumption of a fixed number of raters $J$. A varying number of raters or missing data should therefore be handled within the calculation of matrix $\mathbf{S}$ of possible classifications, with respect to the hierarchy of the categories. As previously, we again introduce the $I \times C$-matrix $\mathbf{J}'$ with the elements $j_{ic}$ representing the number of raters that could have classified subject $i$ into category $c$, *irrespective* of the hierarchy of the categories. This means that $s_{ic}$ is only equal to $j_{ic}$ in the case that $c$ is a main category that is available under all circumstances to raters. In other words: $j_{ic}$ represents the number of possible classifications of subject $i$ into category $c$ without prior knowledge of the other categories the raters have selected (in contrast, this knowledge is definitely required to calculate the matrix $\mathbf{S}$). Hence, matrix $\mathbf{J}'$ is what we need to adjust the denominator of Eq. 13. The scaling factors $\phi_c$ adjusted for a varying number of raters are defined as:

$$\phi_c = \frac{\sum_i s_{ic}}{\sum_i j_{ic}},$$

If the number of raters only varies over subjects (and not over categories), matrix $\mathbf{J}'$s can be replaced by vector $\mathbf{j} = (j_1, j_2, \ldots, j_I)^T$ with $j_i$ defined as the number of raters who classified subject $i$; the adapted $\kappa$ statistic appears by changing matrix $\mathbf{S}$ and the scaling factors $\phi_c$'s accordingly.

## Assumptions, premises, and paradoxes of the proposed measure

Although kappa coefficients are widely used to measure inter-rater agreement, scholars have pointed out that these coefficients are not free from paradoxes and can sometimes yield unexpected results (Warrens, 2010; Gwet, 2008; Feinstein & Cicchetti, 1990). Since the proposed measure generalizes Fleiss' kappa, it inevitably inherits these issues–and even introduces new ones.

In this section, we provide an overview of the assumptions, premises, and paradoxes associated with the proposed $\kappa$-statistic. The subsections serve as a checklist for assessing whether the measure is appropriate for specific research data. The first two subsections address aspects unique to the proposed generalization of Fleiss' kappa, while the latter two focus on criticisms inherited from Fleiss' kappa. Throughout this section, we assume unweighted and non-hierarchical categories.

### Both selecting or not-selecting a category is seen equally as agreement

The original Fleiss' kappa assumes mutually exclusive categories, meaning that two raters $a$ and $b$ can only agree on a subject $i$ if they both classified the subject into the same category $c$, and there are $\binom{x_{ic}}{2}$ such agreeing rater pairs (see the formulas in Section "Fleiss' kappa"). Everything else can be regarded as a disagreement. When a subject can be classi-fied into multiple categories by the same rater, this no longer holds. Indeed, when raters $a$ and $b$ do not select category $c$ for subject $i$, they also agree that from all $C$ categories that can be selected, category $c$ should not be[3]. So the number of agreeing pairs is the sum of $\binom{x_{ic}}{2}$ and $\binom{J-x_{ic}}{2}$, meaning that the agreement on not classifying subject $i$ into category $c$, is valued equally as the agreement on an actual classification of subject $i$ into category $c$ by both raters $a$ and $b$ (see the formulas of the $Po_c$'s and $Pe_c$'s in Equations 5 & 6). This is a philosophical premise of this proposed $\kappa$ statistic, and every user should consider whether this premise is appropriate in a specific context. If the proposed $\kappa$ statistic is used with mutually exclusive, equally weighted, and non-hierarchical categories, Theorem 1 shows that all these terms of agreement on non-classification cancel out.

## Insensitivity to unused categories and the 'always-selected category'-paradox

A valuable feature of the proposed measure is its handling of unused categories. When a category $c$ is not selected by any rater for any subject, i.e., it is not used at all, it is easy to show that $Po_c = Pe_c = 1$, leading to $Po_c - Pe_c = 1 - Pe_c = 0$, resulting in zero contribution to the proposed kappa-statistic (see Equations 7 or 14). Therefore, unused categories do not play any role in the calculation of the proposed $\kappa$ static. This insensitivity from unused categories is often desirable in most research contexts, ensuring the $\kappa$ statistic can not be inflated by simply adding unchosen categories.

On the other hand, it also leads to what we call the 'always-selected category' paradox. From Section "Both selecting or not-selecting a category is seen equally as agreement" we know that selecting or not selecting a category is seen equally as agreement, as such the symmetric case of an unused category arises when a category is used by all raters for all subjects, i.e., the category was always selected. Such 'always-selected category' $c$ leads again to $Po_c = Pe_c = 1$, and so zero contribution to the proposed measure. Although we would expect a category that is chosen by any rater for every subject to have a positive impact on the $\kappa$ statistic, the actual impact is non-existent.

In Table 1, three examples are given of the phenomenon with four raters ($J = 4$), classifying ten subjects ($I = 10$) in 3 categories ($C = 3$). The calculations of these examples are also included as an Excel sheet in the Supplementary material. The first part of Table 1 contains the matrix $\mathbf{X}$ for each example, so each cell $x_{ic}$ is the number of raters that classified subject $i$ in category $c$. In the first example, categories 1 and 2 are unused: not any of the raters used them for at least one subject; while there is disagreement about category 3 that was selected for all subjects by two of the four raters. The second example shows disagreement among all three categories. The third example shows complete agreement for categories 1 and 2, and the same disagreement as the previous examples in category 3. Due to the insensitivity of the proposed measure to non-used categories and always-selected categories, all the examples have a $\kappa = -0.333$, the lowest possible value in this setting with four raters (see Section "Range of possible values"); while intuitively, we would at least expect a much larger value for Example 3, that demonstrated maximal agreement on two categories.

When the 'always-selected category' paradox arises in a certain category $c$, a way to circumvent the paradox is to set $Pe_c = 0$, as such a maximal, positive contribution of 1 will be added to both the numerator as the denominator in the calculation of the $\kappa$ statistic for the always-selected category. This 'always-selected category'-correction leads in Example 3 of Table 1 to $\kappa = 0.733$, clearly distinguishing it from the other examples. Depending on the research context, the correction can also be applied to unused categories, if appropriate.

## The kappa paradox: High observed agreement *Po*, low kappa

One paradox that affects all kappa like measures arises when both the observed agreement $Po$ and the expected chance agreement $Pe$ are high: the correction process embodied in kappa's formula (Eq. 1) can return a relatively low or even negative value of $\kappa$, while the observed agreement $Po$ is high. This is sometimes called the 'kappa paradox' in the literature (Derksen et al., 2024). The proposed measure does not directly measure $Po$ and $Pe$, but sums the $Po_c - Pe_c$ and $1 - Pe_c$ over the categories. As such, the paradox in the proposed measure might manifest itself in a much more disguised way, namely within the contribution of a category to $\kappa$. Therefore, it is always advisable to calculate the $\kappa_c$'s and check them (see Eq. 3).

## The prevalence paradox

Another tightly related paradox is known as the prevalence paradox: it can be shown that the probabilities $\frac{\sum_i x_{ic}}{IJ}$ produce higher $\kappa$ values when they are more balanced, i.e., when all categories are used about equally often and no particularly common categories exist (Warrens, 2010). According to Gwet (2012), these probabilities are not suited to correctly measure the expected chance agreement $Pe$. All ratings for each category are used in the calculation of $Pe$, but as we want to say something about the expected *chance* agreement, this philosophically implies we treat all these ratings as if they were all assigned randomly, which, according to Gwet (2012), is an unacceptable premise. Kraemer et al. (2002) disagree with Gwet's view, saying that 'it is well known that it is very difficult to achieve high reliability of any measure in a very homogeneous population *(of subjects, ed.)*' (p. 2114). Again, the prevalence paradox can arise in the proposed measure only within the contribution of a category to the value of $\kappa$: a category that was selected almost always by all raters (or the symmetric case of an almost unused category), can yield a low $k_c$ with only a few disagreements. An example is presented and explained in the first worked-out example (Section "Worked-out examples").

[3] In fact, the calculation of $\kappa_c$ (Eq. 3) is equal to the calculation of the Fleiss' kappa with two categories: 'selected category $c$' and 'not-selected category $c$.'

**Table 1** Example of the effects of the 'always-selected category'-paradox with $I = 10$, $J = 4$ and $C = 3$

| | Example 1 | | | Example 2 | | | Example 3 | | |
|---|---|---|---|---|---|---|---|---|---|
| | Cat 1 | Cat 2 | Cat 3 | Cat 1 | Cat 2 | Cat 3 | Cat 1 | Cat 2 | Cat 3 |
| Subject 1 | 0 | 0 | 2 | 2 | 2 | 2 | 4 | 4 | 2 |
| Subject 2 | 0 | 0 | 2 | 2 | 2 | 2 | 4 | 4 | 2 |
| Subject 3 | 0 | 0 | 2 | 2 | 2 | 2 | 4 | 4 | 2 |
| Subject 4 | 0 | 0 | 2 | 2 | 2 | 2 | 4 | 4 | 2 |
| Subject 5 | 0 | 0 | 2 | 2 | 2 | 2 | 4 | 4 | 2 |
| Subject 6 | 0 | 0 | 2 | 2 | 2 | 2 | 4 | 4 | 2 |
| Subject 7 | 0 | 0 | 2 | 2 | 2 | 2 | 4 | 4 | 2 |
| Subject 8 | 0 | 0 | 2 | 2 | 2 | 2 | 4 | 4 | 2 |
| Subject 9 | 0 | 0 | 2 | 2 | 2 | 2 | 4 | 4 | 2 |
| Subject 10 | 0 | 0 | 2 | 2 | 2 | 2 | 4 | 4 | 2 |
| $Po_c$ | 1 | 1 | 0.333 | 0.333 | 0.333 | 0.333 | 1 | 1 | 0.333 |
| $Pe_c$ | 1 | 1 | 0.5 | 0.5 | 0.5 | 0.5 | 1 | 1 | 0.5 |
| $Po_c - Pe_c$ | 0 | 0 | -0.167 | -0.167 | -0.167 | -0.167 | 0 | 0 | -0.167 |
| $1 - Pe_c$ | 0 | 0 | 0.5 | 0.5 | 0.5 | 0.5 | 0 | 0 | 0.5 |
| $\kappa_c$ | NaN | NaN | -0.333 | -0.333 | -0.333 | -0.333 | NaN | NaN | -0.333 |
| $\kappa$ | **-0.333** | | | **-0.333** | | | **-0.333** | | |

## Range of possible values

Contrary to what is frequently assumed, the original Fleiss' kappa does not have a range from $[-1, 1]$ when more than two raters are involved (Vanbelle, 2009). Remember that in the original Fleiss' kappa, all raters select one category for each subject.

In this section, we first examine the range of the original Fleiss' kappa and then establish its connection to our proposed $\kappa$ measure. Throughout, we assume unweighted, non-hierarchical categories.

### Fleiss' kappa

When the number of categories equals the number of raters ($J = C$), the maximum possible disagreement in the original Fleiss' kappa occurs when each rater selects a different category for every subject. In this scenario, $x_{ic} = 1$ for all subject-category combinations $(i, c)$. Consequently, the observed proportion of agreement $P_o$ is 0, while the expected proportion of agreement $P_e$ is $\frac{1}{J}$, leading to $\kappa = -\frac{1}{J-1}$. These calculations can be verified using the equations from Section "Fleiss' kappa".

The situation becomes more complex when $J < C$. In this case, maximum disagreement still occurs when each rater selects a different category for every subject. However, since there are more categories than raters, at least one category will remain unselected for each subject, resulting in $x_{ic}$ values of either 1 or 0. While $P_o$ remains 0, the expected agreement $P_e$ equals $\frac{1}{J}$ only if the same categories are consistently unselected across all subjects. If, instead, the unselected categories vary between subjects – that is, if not all subjects have the same subset of unselected categories – then $P_e$ will be slightly lower than $\frac{1}{J}$, leading to a slightly higher $\kappa$ value than $-\frac{1}{J-1}$.

When the number of raters exceeds the number of categories ($J > C$), the Pigeonhole principle (Brualdi, 2010) guarantees that at least one category will be assigned by multiple raters for each subject, resulting in a certain degree of 'agreement by design.' Consequently, the minimum possible kappa value in this scenario is greater than $-\frac{1}{J-1}$, as the maximum level of disagreement observed when $J \leq C$ is no longer achievable.

In all cases, Fleiss' kappa reaches its maximum value of 1 when $P_o = 1$ and $P_e \neq 1$. Thus, the range of the original Fleiss' kappa is $\left[-\frac{1}{J-1}, 1\right]$, with the minimum value attainable only when $J \leq C$.

## Our proposed kappa statistic

As a generalization of Fleiss' kappa, our proposed kappa statistic exhibits the same range of possible values, $\left[-\frac{1}{J-1}, 1\right]$. Naturally, as established in Theorem 1, the same cases of maximum disagreement can occur, yielding identical $\kappa$ values.

Recall from Section "Both selecting or not-selecting a category is seen equally as agreement" that selecting a category is considered equally indicative of agreement as not selecting it. Consequently, maximum disagreement can also occur with an even number of raters $J$ if, for every subject and every category, half of the raters select the category while the other half do not. In this case, $Po_c = \frac{J-2}{2J-2}$ (Eq. 5) and $Pe_c = \frac{1}{2}$ (Eq. 6), leading again to $\kappa = -\frac{1}{J-1}$. This corresponds to Example 2 in Table 1. For an odd number of raters $J$, only settings similar to those described in the previous paragraph will result in the minimum $\kappa$ value.

## Interpreting & benchmarking the new measure: the 'Interval Membership Probability' (IMP) method

How should we interpret the new measure? When is the extent of agreement 'good enough'? The most widely adopted benchmark scale is proposed by Landis and Koch (1977) and is shown in Table 2. With over 92,000 registered citations (and counting) on Google Scholar, it has been prevalent among many researchers for a long time. However, the theoretical underpinnings of the benchmark scale are scant: (1) the benchmark scale was introduced for Cohen's kappa on $2 \times 2$ agreement tables but used in practice with all chance-corrected agreement coefficients, (2) the benchmark scale is based on personal experience of Landis and Koch with the Cohen's $\kappa$, with no evidence to support it, and most importantly: (3) it ignores the experimental conditions from which the calculated coefficient originates. As a result, directly comparing any $\kappa$ value against a benchmark scale can be misleading, as it fails to account for differences in the number of subjects, categories, and raters across studies. The previous section illustrates this issue clearly, showing that even the minimum possible kappa value depends on the specific number of raters and categories involved.

**Table 2** Benchmark scale of Landis and Koch (1977)

| $\kappa$-statistic | Interpretation |
|---|---|
| 0.81 to 1.00 | Almost perfect agreement |
| 0.61 to 0.80 | Substantial agreement |
| 0.41 to 0.60 | Moderate agreement |
| 0.21 to 0.40 | Fair agreement |
| 0.00 to 0.20 | Slight agreement |
| $< 0$ | Poor agreement |

Gwet (2012) developed a method based on the probability that a $\kappa$ statistic falls into each agreement level of any benchmark scale; arguing that the used benchmark scale does not matter, as long as benchmarking entails 'A statement saying how confident we are that the extent of agreement among raters reaches a certain agreement level. Good benchmarking will be statistical, not deterministic' (p. 173). The method was later called the 'Interval Membership Probability' (IMP) method by Vanacore and Pellegrino (2022). As these probabilities (IMPs) are based on the standard error of the agreement coefficients, the method allows for a comparison across different experimental conditions. Well-designed studies will have a lower standard error and thus higher benchmarking probabilities. At the same time, poorly designed experiments are prevented from producing 'almost perfect' agreements solely based on an imprecisely estimated $\kappa$ statistic.

For established agreement coefficients (Martín Andrés et al., 2025), standard error formulas have been derived that can be used for calculating the membership probabilities (IMPs) over the standard normal distribution, as done in Gwet (2012). As the large-sample variance of our proposed measure still needs to be determined, bootstrap resampling can be used to determine the IMPs. As such, we also do not make any distributional assumption.

This bootstrapped 'Interval Membership Probability' (IMP) method applied to the proposed measure consists of three steps:

1. Use bootstrap resampling to repeatedly calculate the proposed measure. In general, 10,000 bootstrap samples are advised. The bootstrap samples consist of random samples of the subjects (and their classifications) with replacements (so $n = I$ for each bootstrap sample, but the subjects can be included repeatedly). More mathematically, you resample with replacements the rows of matrix $\mathbf{X}$ (Vanbelle & Albert, 2008).
2. Calculate the probability that the extent of agreement falls into each category of the benchmark scale. This probability is the IMP. With the scale of Landis and Koch and 10,000 bootstrap samples, the IMP for 'Almost perfect' can be calculated by looking at how many of these samples returned a $\tilde{\kappa} > 0.8$.
3. Compute the cumulative IMP for each category of the benchmarking scale, starting from the highest category to the lowest.
4. The final level of agreement is the category with the smallest cumulative IMP exceeding $(1 - \alpha\%)$, with $\alpha$ being the chosen alpha level.

R-code to execute the method is provided as Supplementary material. In the worked-out examples, we also show how the IMP method works in practice.

## Worked-out examples

In this section, we apply our proposed $\kappa$ statistic and the appropriate other methods from the introduction to two applications: one on the assessment of a mathematics exam for which our proposed $\kappa$ statistic was initially developed, the other is an example from Mezzich et al. (1981) in which 30 child psychiatrists diagnose patients into multiple psychiatric disorders. Of course, there are plenty of other applications of the measure as well (e.g., coding qualitative data in multiple categories, behavioral observations, peer review).

The R code, data files, and comprehensive Excel spreadsheets accompany this article as Supplementary materials (also available at https://osf.io/q5nft/) to facilitate understanding of the calculations in the worked-out examples. The spreadsheets allow users to modify the classifications in the examples, thereby demonstrating their impact on the proposed kappa measure. Additionally, the spreadsheets can be easily adapted to the reader's research data. The R-script contains ready-to-use functions to calculate the proposed measure.

### Assessing mathematics exams

#### Context

The proposed $\kappa$ statistic was initially developed to measure the inter-rater agreement of multiple teachers assessing students with a new assessment method (Moons et al., 2025) for handwritten high-stakes mathematics exams called 'checkbox grading.' The method allows exam designers to preset a list of feedback items with partial scores for each question, so that teachers can just tick the items (= categories) relevant to a student's answer. Hierarchical dependencies between items can be set, so items can be shown, disabled, or adapted whenever a previous item is ticked, implying that teachers must follow the preset point-by-point feedback items from top to bottom. This adaptive grading approach resembles a flow chart that automatically determines the grade. Moreover, checking the items that are relevant to a student's answer leads at the same time lead to: (1) a deep insight into how the grade was obtained for both student feedback (Moons et al., 2024) as well as the exam designers, and (2) a straightforward way to do correction work with multiple teachers where personal interpretations are avoided as much as possible.

An example of checkbox grading is given in Fig. 2. With this drawing question, a student can gain a maximum score of 3 points. If point $A$ is drawn correctly (1$^{\text{st}}$ bullet), the student gains 1 point; the correct drawing of point $B$ (3$^{\text{rd}}$ bullet) is worth 1.5 points. The 2$^{\text{nd}}$ bullet does not change the score but shows teachers that the presence of auxiliary lines is perfectly fine. The last two feedback items, bullets 4 and 5, can only be selected if items 1 and 3 were selected. As the drawing of the line $AB$ implies the drawing of the line segment $[AB]$, the 5$^{\text{th}}$ bullet can only be selected if the 4$^{\text{th}}$ was. This is a clear example of hierarchical items (= categories).

During the project, one of the main research questions concerned the inter-rater agreement of this new assessment method with multiple assessors (Moons et al., 2025). The traditional measures for inter-rater reliability such as intraclass correlations, fell short because these can only measure the agreement between teachers on grades, while the method also provides feedback to students. Hence, it is not enough to agree on grades; the resulting feedback to the students must also be as equal as possible. Score agreement by no means guarantees agreement on feedback items, which is especially clear for feedback items not influencing the score (e.g., bullet 2 in the example). Other examples can be given as well: in Fig. 2, 2.5 points can be obtained by solely drawing points $A$ and $B$ correctly (only bullets 1 and 3 apply, possibly bullet 2) or by drawing the line $AB$ correctly (all bullets apply, possibly bullet 2). Conversely, the inverse is true: agreement on feedback items implies score agreement.

Our proposed $\kappa$ statistic with hierarchical categories does meet all requirements:

- It will assess the agreement of the raters in selecting multiple feedback items (= categories) for each student (= subjects)
- These items are hierarchical: the selectability of some items depends on the selection of other items
- Score agreement can naturally be measured by weighing the items according to their partial scores.

#### Example

We start with a worked-out example, in which our proposed $\kappa$ statistic is calculated step-by-step. We consider three teachers (i.e., the number of raters $J$ equals 3) assessing six students' solutions (i.e., the number of subjects $I$ equals 6) on the question in Fig. 2. The teachers classified every student's solution into the five checkboxes/feedback items (i.e., the number of categories $C$ equals 5). The classifications by the three teachers of the six students' answers can be found in Table 3. Although the example consists of a simple ques-tion, the three teachers (raters) did sometimes select different items (categories) for the students' solutions (subjects).

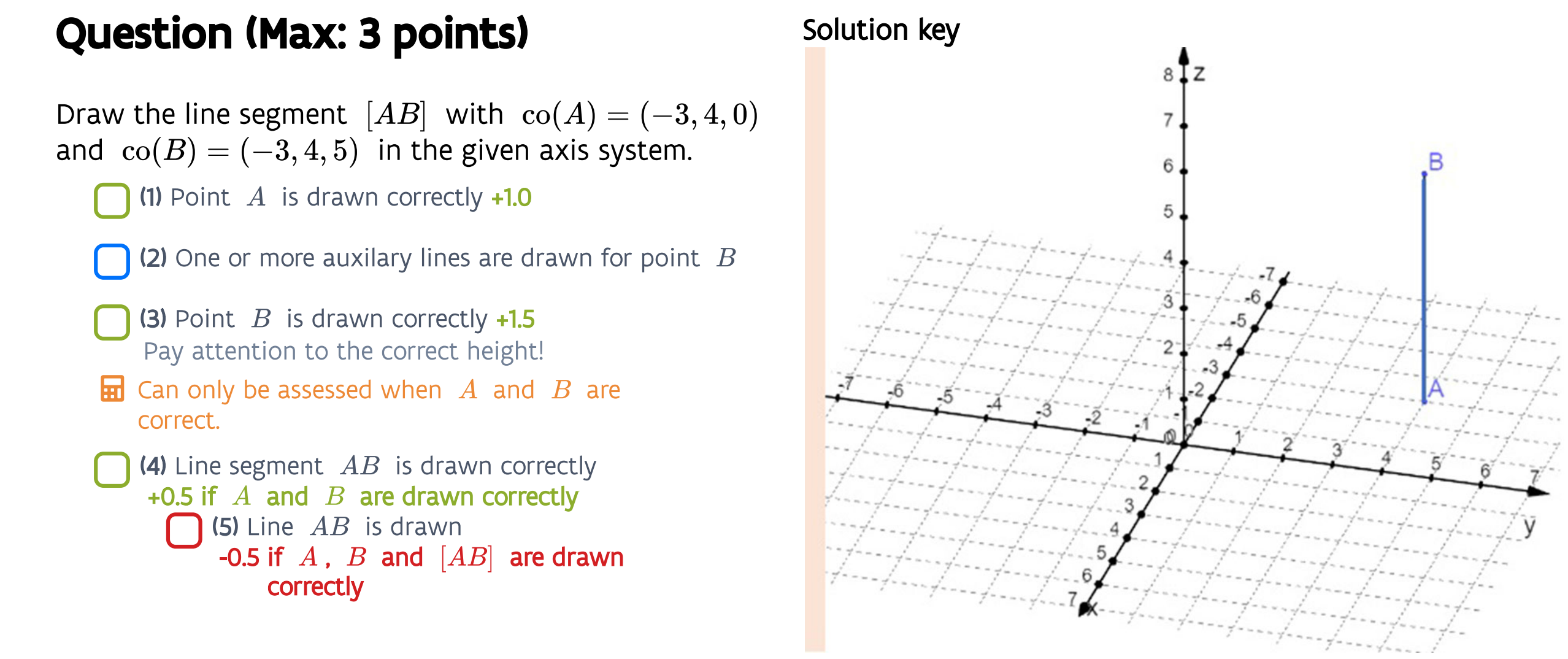

**Fig. 2** Example question with checkbox grading

### Specification of the weight vector w

We start by specifying the weights of the vector **w**. The associated scores for each item will evidently play a crucial role in defining these. However, note that in Fig. 2 the second (blue) item does not influence the final grade on the question. If our weights only represent the associated scores, then $w_2 = 0$; meaning that item 2 would not play any role in the calculation of our kappa-statistic, while the presence/absence of the item changes the feedback a student receives. Hence, instead of using the (absolute value) of the associated score to define the weights, we add the maximum absolute value of the associated scores over all items. This means that the weights will be defined based on $|\text{score}_c| + \max_{k=1}^{C}\{|\text{score}_k|\}$. To get weights between 0 and 1, we divide this sum by the doubled maximum associated score over all items:

$$w_c = \frac{|\text{score}_c| + \max_{k=1}^{C}\{|\text{score}_k|\}}{2 \cdot \max_{k=1}^{C}\{|\text{score}_k|\}}. \quad (15)$$

These weights have a nice interpretation: the minimum weight is always 0.5, accounting for the (non-)selection of the item, everything between 0.5 and 1 depends on the (absolute value of) the associated score of the item. As such, items that do not influence the final score, will have a weight of 0.5, while items with the maximum (absolute value of the) associated score will have weight 1. These weights do not sum to 1, considering their interpretation is more intuitive this way. Based on Eq. 15, the calculated weights for the example are given in Table 4.

### Determining the matrix of possible classifications S and scale factors $\phi_c$ based on the hierarchical structure of the categories

We see that the first three items are all main categories: there are no conditions for (not) selecting them, so $s_{i1} = s_{i2} = s_{i3} = J = 3$ for every student $i$. For a possible classification into item 4, item 1 and item 3 must be selected first; for example, student 6 has only the third teacher selecting these, so $s_{6,4} = 1$. Item 5 can only be selected if item 4 was selected so $s_{i5} = x_{i4}$; for example, student 1 has 2 classifications for item 4 (teacher 1 & teacher 3), so $s_{1,5} = 2$. Matrix **S** can be found in Table 5.

**Table 3** Assessments by three teachers of six student's answers on the example question

| | Teacher 1 | | | | | | Teacher 2 | | | | | | Teacher 3 | | | | | |
|---|---|---|---|---|---|---|---|---|---|---|---|---|---|---|---|---|---|---|
| | S1 | S2 | S3 | S4 | S5 | S6 | S1 | S2 | S3 | S4 | S5 | S6 | S1 | S2 | S3 | S4 | S5 | S6 |
| (1) | X | X | X | X | X | X | X | | X | X | X | X | X | | X | X | X | X |
| (2) | | | X | X | X | X | | | X | X | X | X | | | | X | X | X |
| (3) | X | | | X | X | | X | | | X | X | | X | | | X | X | X |
| (4) | X | | | X | X | | | | | X | X | | X | | | X | X | X |
| (5) | | | | | X | | | | | | X | | | | | | X | |
| Score | 3 | 1 | 1 | 3 | 2.5 | 1 | 2.5 | 0 | 1 | 3 | 2.5 | 1 | 3 | 0 | 1 | 3 | 2.5 | 3 |

**Table 4** Specification of the weight vector **w**

| Item | | (1) | (2) | (3) | (4) | (5) |
|---|---|---|---|---|---|---|
| $\lvert \text{score}_c \rvert$ | (associated score) | 1 | 0 | 1.5 | 0.5 | 0.5 |
| $\max_{k=1}^{C}\{\lvert \text{score}_k \rvert\}$ | (selection) | 1.5 | 1.5 | 1.5 | 1.5 | 1.5 |
| Sum | | 2.5 | 1.5 | 3 | 2 | 2 |
| **Weight $w_c$** | | **0.833** | **0.5** | **1** | **0.667** | **0.667** |

The scale factors $\phi_c$ can be found by applying formula Eq. 13: for each category $c$, loop over all subject $i$ and take the sum of the $s_{ic}$'s (sum up the columns of Table 5), and divide this sum by $IJ = 6 \cdot 3 = 18$.

### Calculating $Po_c$ and $Pe_c$

We give the full calculation of $Po_1$ and $Pe_1$ in this paragraph. The other $Po_c$'s and $Pe_c$'s can be calculated in a similar way. The required $s_{i1}$ values were already calculated in the previous step; we still need to count how many times item 1 was selected for each student $i$ to get the $x_{i1}$ values; the results can be found in Table 6.

Next, we calculate $Po_1$ based on formula Eq. 11:

$$Po_1 = \frac{[3 \cdot 2 + 0 \cdot -1] + [1 \cdot 0 + 2 \cdot 1] + [3 \cdot 2 + 0 \cdot -1] + [3 \cdot 2 + 0 \cdot -1] + [3 \cdot 2 + 0 \cdot -1] + [3 \cdot 2 + 0 \cdot -1]}{3 \cdot 2 + 3 \cdot 2 + 3 \cdot 2 + 3 \cdot 2 + 3 \cdot 2 + 3 \cdot 2}$$
$$= 0.889,$$

For the computation of $Pe_1$, we use Eq. 12:

$$Pe_1 = 2 \cdot \left(\frac{3+1+3+3+3+3}{3+3+3+3+3+3}\right)^2 - 2 \cdot \left(\frac{3+1+3+3+3+3}{3+3+3+3+3+3}\right) + 1$$
$$= 0.802,$$

**Table 5** Determining the matrix of possible classifications **S** and scale factors $\phi_c$

| | $s_{i1}$ | $s_{i2}$ | $s_{i3}$ | $s_{i4}$ | $s_{i5}$ |
|---|---|---|---|---|---|
| $s_{1c}$ | 3 | 3 | 3 | 3 | 2 |
| $s_{2c}$ | 3 | 3 | 3 | 0 | 0 |
| $s_{3c}$ | 3 | 3 | 3 | 0 | 0 |
| $s_{4c}$ | 3 | 3 | 3 | 3 | 3 |
| $s_{5c}$ | 3 | 3 | 3 | 3 | 3 |
| $s_{6c}$ | 3 | 3 | 3 | 1 | 1 |
| Sum | 18 | 18 | 18 | 10 | 9 |
| Scale factors | $\phi_1$ | $\phi_2$ | $\phi_3$ | $\phi_4$ | $\phi_5$ |
| | 1 | 1 | 1 | 0.556 | 0.5 |

Although not necessary for the calculation of our proposed $\kappa$ statistic, it is possible to calculate the partial $\kappa_c$ to have an indication of the agreement within each item. For item 1, this becomes (see formula Eq. 3):

$$\kappa_1 = \frac{Po_1 - Pe_1}{1 - Pe_1} = \frac{0.889 - 0.802}{1 - 0.802} = 0.438.$$

Although item 1 was selected for most students (only teachers 2 and 3 did not select it for student 2), we get a relatively low $\kappa_1$ value. How can this be explained? Item 1 was chosen for almost all students by almost all teachers, leading to a high agreement by chance $Pe_1 (= 0.802)$. This means that without even looking at a student's solution, there is a high probability that a teacher selects item 1. The fact that student 2 has two non-classifications for item 1 while teacher 1 did select item 1 for this student leads, therefore, leads to a pretty severe penalization in the partial kappa $\kappa_1$. This is a concrete example of the 'prevalence paradox' described in Section "The prevalence paradox".

The other $Po_c$'s and $Pe_c$'s can be calculated analogously. The result can be found in Table 7.

### Calculation of the kappa statistic

With the specification of weight vector **w**, and the computation of the scale factors $\phi_c$, the 'beyond-chance' $Po_c - Pe_c$ and the 'beyond-chance in case of perfectly agreeing raters'

**Table 6** Determining the $x_{i1}$'s and $s_{i1}$'s

| Student | S1 | S2 | S3 | S4 | S5 | S6 |
|---|---|---|---|---|---|---|
| $x_{i1}$ | 3 | 1 | 3 | 3 | 3 | 3 |
| $s_{i1}$ | 3 | 3 | 3 | 3 | 3 | 3 |

$1 - Pe_c$, we are ready to calculate the kappa statistic for the hierarchical case (see Eq. 14):

$$\kappa = \frac{0.833 \cdot 1 \cdot 0.086 + 0.5 \cdot 1 \cdot 0.364 + 1 \cdot 1 \cdot 0.383 + 0.667 \cdot 0.556 \cdot (-0.042) + 0.667 \cdot 0.5 \cdot 0.444}{0.833 \cdot 1 \cdot 0.198 + 0.5 \cdot 1 \cdot 0.475 + 1 \cdot 1 \cdot 0.494 + 0.667 \cdot 0.556 \cdot 0.180 + 0.667 \cdot 0.5 \cdot 0.444}$$
$$= 0.692.$$

### Interpreting the kappa statistic with the Interval Membership Probability method (IMP)

We get a relatively high $\kappa$ value that would be labeled by the benchmark scale of Landis and Koch (1977) as 'Substantial' agreement. Let's see what the IMP method tells with $\alpha = 0.05$. Therefore, we calculate the membership probabilities (IMP) and cumulative IMP for each benchmark range of Landis and Koch by using the empirical probabilities of 10,000 bootstrap samples using the provided R-code in the Supplementary materials, see Table 8 for the results.

Based on Table 8, we can claim almost substantial agreement with a probability of only 88.54%. Meaning that of the 10,000 bootstrap samples, 8854 resulted in a kappa value greater than or equal to 0.61. Based on the cumulative IMPs, we are 95.73% certain the level of agreement was fair. As that category is the first to exceed a cumulative IMP of 95%, it is the final level of agreement by the IMP method. The much lower agreement level than direct comparison can be explained by the low number of raters ($J = 3$), subjects ($I = 6$) and categories ($C = 5$), leading to a large variability of empirical $\kappa$ values.

### Comparison with other methods

We also calculated this example through the other methods described in the Introduction. Averaging/pooling Cohen's kappas is not possible, as we have more than two raters. The proportional overlap method is possible and returns $\kappa = 0.602$. However, the method is based on some questionable premises in this context: (1) it assumes all items are equally weighted (so there is no correction for the associated scores), (2) it assumes all categories are always available to all raters (so the hierarchy of the items is ignored). Besides, the method fails to measure potential observed agreement for student 2 as teacher 2 and 3 did not select any category for this student, no proportional overlaps can be calculated. Problems (1) and (2) also occur with the chance-corrected intraclass correlations that return a $\kappa$ value of 0.379. The problem of failing to measure potential observed agreement for student 2 emerges in another guise, while the proportional overlap method leaves student 2 out of the calculation of $Po$, the chance-corrected intraclass correlations do include student 2 with an intraclass correlation coefficient of almost zero, pulling down the $Po$ value in an unacceptable way. While our proposed $\kappa$ statistic entails the philosophical premise that two raters not selecting category $c$ is equally valued in terms of agreement as two raters who do select category $c$; these examples show that the opposite – completely excluding agreement in non-selections – can also lead to unsatisfactory results. Finally, the calculation of chance-corrected rank correlations is not relevant in this context, as raters do not make ordered classifications in checkbox grading.

**Table 7** $Po_c$, $Pe_c$, $Po - Pe$, $1 - Pe$ and partial kappa $\kappa_c$ for every item (=category)

| Items | (1) | (2) | (3) | (4) | (5) |
|---|---|---|---|---|---|
| $Po_c$ | 0.889 | 0.889 | 0.889 | 0.778 | 1.00 |
| $Pe_c$ | 0.802 | 0.525 | 0.506 | 0.820 | 0.556 |
| $Po_c - Pe_c$ | 0.086 | 0.364 | 0.383 | -0.042 | 0.444 |
| $1 - Pe_c$ | 0.198 | 0.475 | 0.494 | 0.180 | 0.444 |
| $\kappa_c$ | 0.438 | 0.766 | 0.775 | -0.235 | 1.00 |

## Diagnosing psychiatric cases

We now revisit an example from Mezzich et al. (1981). It consists of a diagnostic exercise in which 30 child psychiatrists made independent diagnoses of 27 child psychiatric cases. Each psychiatrist rated three cases, and each case turned out to be rated by three or four psychiatrists upon completion of the study. Table 9 shows the 90 multiple diagnostic formulations. Each diagnostic formulation presented was composed of up to three from the twenty broad diagnostic categories taken from Axis I (clinical psychiatric syndromes) of the American Psychiatric Association's Diagnostic and Statistical Manual of Mental Disorders (DSM-III). We are well aware that DSM-III is outdated (American Psychiatric Association, 2022), but the example remains excellent as it can be contrasted with the other measures in the literature.

**Table 8** Using the IMP method to analyse the 'mathematics exam' example with 10,000 bootstrap samples

| Benchmark range | Agreement level | IMP | Cumulative IMP |
|---|---|---|---|
| 0.81 to 1.00 | Almost perfect | 0.7712 | 0.7712 |
| 0.61 to 0.80 | Substantial | 0.1142 | 0.8854 |
| 0.41 to 0.60 | Moderate | 0.0561 | 0.9415 |
| 0.21 to 0.40 | Fair | 0.0168 | 0.9583 |
| 0.00 to 0.20 | Slight | 0.0033 | 0.9616 |
| $< 0$ | Poor | 0.0384 | 1.0000 |

We start with the calculation of our proposed $\kappa$ statistic. The example consists of 27 child psychiatric cases (i.e., the number of subjects $I$ equals 27), to be classified into 20 broad diagnostic categories (i.e., the number of categories $C$ equals 20) with a varying number of raters, expressed in vector $\mathbf{j}$ with $j_i = 3$ or $j_i = 4$, depending on the case, see Table 10.

We assume all diagnostic categories are equally important and thus use unweighted categories. Moreover, the diagnostic categories on Axis I have no hierarchy. Hence, we can use the formulas of our proposed kappa statistic with a varying number of raters in the non-hierarchical, unweighted case. First, we calculate matrix $\mathbf{X}$ by counting how many times a diagnostic category $c$ appeared for a subject $i$ (e.g., $x_{1,1} = 0$, $x_{12,1} = 2$, $x_{6,10} = 3, \ldots$). Next, we combine the $x_{ic}$'s and the $j_i$'s to determine the $Po_c$'s (Eq. 9) and the $Pe_c$'s (Eq. 10). As an example, we calculate $Po_1$ and $Pe_1$:

$$Po_1 = \frac{9\left[2\cdot 0^2 - 2\cdot 4\cdot 0 + 4^2 - 4\right] + 16\left[2\cdot 0^2 - 2\cdot 3\cdot 0 + 3^2 - 3\right] + 1\left[2\cdot 1^2 - 2\cdot 3\cdot 1 + 3^2 - 3\right] + 1\left[2\cdot 2^2 - 2\cdot 3\cdot 2 + 3^2 - 3\right]}{9\,[4(4-1)] + 18\,[3(3-1)]}$$
$$= 0.963$$
$$Pe_1 = 2\left(\frac{9\cdot 0 + 16\cdot 0 + 1\cdot 1 + 1\cdot 2}{9\cdot 4 + 18\cdot 3}\right)^2 - 2\left(\frac{9\cdot 0 + 16\cdot 0 + 1\cdot 1 + 1\cdot 2}{9\cdot 4 + 18\cdot 3}\right) + 1$$
$$= 0.936.$$

The other calculations can be found in Table 11.

Note that $\kappa_2$, $\kappa_4$, $\kappa_6$ and $\kappa_{19}$ equal NaN, due to a division by zero, ensuring the kappa statistic is not inflated by unused categories (see Section "Insensitivity to unused categories and the 'always-selected category'-paradox"), we get from formula Eq. 7:

$$\kappa = \frac{\sum_c (Po_c - Pe_c)}{\sum_c (1 - Pe_c)} = 0.375.$$

We get a relatively low kappa value, which should not come as a surprise: Table 9 shows that the various psychiatrists diverge rather vehemently in their diagnoses. Using the IMP method (Section "Interpreting & benchmarking the new measure: the 'Interval Membership Probability' (IMP) method"), we have 99.4% confidence that this value falls in the 'Fair' agreement level of Landis and Koch (1977; see R-script for calculations).

The proposed $\kappa$ statistic yields a higher value than the proportional overlap method ($\kappa = 0.27$), but is almost equal to the chance-corrected intraclass correlation method ($\kappa = 0.35$) and the rank correlation method ($\kappa = 0.38$). The rank correlation method is calculated by considering the first diagnosis in Table 9 as the primary diagnosis (rank = 1), the second gets rank 2, etc.

## Future research

The story of the proposed $\kappa$ statistic is not finished by publishing this paper. Indeed, more can be told about the proposed measure. Based on De Vries et al. (2008), we envision publishing the simulation study to show that our proposed kappa statistic exhibits smaller root-mean-square errors than taking a weighted average of Fleiss' kappas. Moreover, the large-sample variance of the proposed $\kappa$ statistic still needs to be determined. An expression for the variance would enable statistical inference using the measure without bootstrapping. It especially paves the way for performing robust power analysis: researchers wishing to set up an experiment in which raters classify subjects into one or more categories would be able to calculate in advance the number of raters and subjects required to reach a certain confidence level.

An additional avenue for further research concerns hierarchical classification structures. While our proposed measure accommodates such structures through the concept of possible classifications ($s_{ic}$), and a worked-out example is provided, future research could explore whether alternative formulations or partial-weighting schemes across levels might enhance interpretability or applicability.

Finally, now that we have established the idea of the proposed $\kappa$ statistic, the same idea may be suitable to create other long-needed measures. For example, the literature on rubrics (Dawson, 2017) lacks a unified way to compare the inter-rater agreement of two rubrics assessing the same phenomenon (e.g., book reviews of students, PhD proposals). Should such a measure exist, it would be possible to compare the impact of including/excluding specific criteria. Such a measure can possibly be constructed by the calculation of the $Po$ and $Pe$ of the Fleiss kappa (or the Krippendorff's

**Table 9** Multiple diagnostic formulations from 27 child psychiatric cases using DSM-III Axis I Broad Categories*

| Cases | Raters 1 | 2 | 3 | 4 |
|---|---|---|---|---|
| 1 | 9, 11 | 11, 9, 14 | 16, 9 | 11, 9 |
| 2 | 16 | 16, 14 | 12 | 14, 5 |
| 3 | 17 | 12 | 7, 8 | 13 |
| 4 | 16, 13 | 13, 16, 14 | 16 | |
| 5 | 7 | 7, 12, 13 | 13 | |
| 6 | 10 | 10 | 10 | |
| 7 | 7, 16 | 13 | 16 | |
| 8 | 1, 14 | 13 | 16, 13 | |
| 9 | 5 | 20 | 13, 14 | |
| 10 | 12, 13, 14 | 12, 14, 13 | 12, 11, 14 | |
| 11 | 13 | 18 | 16 | |
| 12 | 5, 18 | 1, 5, 18 | 1 | |
| 13 | 14, 13 | 14, 7 | 14, 16 | |
| 14 | 11, 16 | 14, 11, 16 | 11, 13 | |
| 15 | 10 | 3, 18 | 10, 11 | |
| 16 | 14, 5 | 5, 16 | 14 | |
| 17 | 12 | 12, 11 | 12 | |
| 18 | 20 | 16 | 16 | |
| 19 | 13 | 14 | 14 | |
| 20 | 9, 14, 10 | 9, 11, 14 | 10, 9 | |
| 21 | 12, 11 | 11, 14 | 11 | |
| 22 | 17 | 12 | 12 | 12, 17, 15 |
| 23 | 16, 13 | 12 | 14 | 13 |
| 24 | 12 | 12 | 16 | 12 |
| 25 | 13 | 20 | 13 | 13 |
| 26 | 13 | 13, 16 | 13 | 16 |
| 27 | 10, 9 | 9, 10 | 9 | 9, 10 |

* 1. Organic mental disorders, 2. Substance use disorders, 3. Schizophrenic and paranoid disorders, 4. Schizoaffective disorders, 5. Affective disorder, 6. Psychoses not elsewhere classified, 7. Anxiety factitious, somatoform and dissociative disorders, 8. Pyschosexual disorder, 9. Mental retardation, 10. Pervasive developmental disorder, 11. Attention deficit disorders, 12. Conduct disorders, 13. Anxiety disorders of childhood or adolescence, 14. Other disorders of childhood or adolescence, speech and stereotyped movement disorders, disorders characteristic of late adolescence, 15. Eating disorders, 16. Reactive disorders not elsewhere classified, 17. Disorders of impulse control not elsewhere classified, 18. Sleep and other disorders, 19. Conditions not attributable to a mental disorder, 20. No diagnosis on Axis I.

alpha, see Gwet, 2012) for groups of criteria assessing the same aspect and weighting them according to the maximum score of the aspect.

## Conclusion

This paper has presented a generalization of Fleiss' kappa, allowing raters to select multiple categories for each subject. Categories can be weighted according to their importance in the research context, and the measure can account for possible hierarchical dependencies between the categories. A crucial assumption of the proposed $\kappa$ statistic is that two raters selecting a specific category for a given subject count equally in agreement as two raters not selecting the category. Other methods, like proportional overlap, chance-corrected intraclass correlations, and chance-corrected rank correlations, do not make this assumption; instead, they ignore the agreement in the non-selection of categories. We have shown that this ignorance can give unexpected and unwanted results depending on the research context. By introducing this generalization of Fleiss' kappa and comparing and contrasting it to the existing comparable methods, we hope to inspire further researchers in need of a chance-corrected inter-rater agreement measure that allows measuring the agreement among several raters classifying subjects into one or more (hierarchical) nominal categories.

**Table 10** Number of psychiatrists (= raters) for each case $i$ (= subject)

| Cases | 1 | 2 | 3 | 4 | 5 | 6 | 7 | 8 | 9 | 10 | 11 | 12 | 13 | 14 | 15 | 16 | 17 | 18 | 19 | 20 | 21 | 22 | 23 | 24 | 25 | 26 | 27 |
|---|---|---|---|---|---|---|---|---|---|---|---|---|---|---|---|---|---|---|---|---|---|---|---|---|---|---|---|
| $j_i$ | 4 | 4 | 4 | 3 | 3 | 3 | 3 | 3 | 3 | 3 | 3 | 3 | 3 | 3 | 3 | 3 | 3 | 3 | 3 | 3 | 3 | 4 | 4 | 4 | 4 | 4 | 4 |

**Table 11** $Po_c$, $Pe_c$, $Po - Pe$, $1 - Pe$ and partial kappa $\kappa_c$ for every diagnostic category

| Diagnostic category | 1 | 2 | 3 | 4 | 5 | 6 | 7 | 8 | 9 | 10 |
|---|---|---|---|---|---|---|---|---|---|---|
| $Po_c$ | 0.963 | 1.000 | 0.981 | 1.000 | 0.917 | 1.000 | 0.917 | 0.972 | 1.000 | 0.935 |
| $Pe_c$ | 0.936 | 1.000 | 0.978 | 1.000 | 0.876 | 1.000 | 0.895 | 0.978 | 0.785 | 0.802 |
| $Po_c - Pe_c$ | 0.027 | 0.000 | 0.003 | 0.000 | 0.041 | 0.000 | 0.022 | -0.006 | 0.215 | 0.133 |
| $1 - Pe_c$ | 0.064 | 0.000 | 0.022 | 0.000 | 0.124 | 0.000 | 0.105 | 0.022 | 0.215 | 0.198 |
| $\kappa_c$ | 0.425 | NaN | 0.157 | NaN | 0.330 | NaN | 0.206 | -0.264 | 1.000 | 0.672 |
| Diagnostic category | 11 | 12 | 13 | 14 | 15 | 16 | 17 | 18 | 19 | 20 |
| $Po_c$ | 0.898 | 0.824 | 0.694 | 0.759 | 0.972 | 0.713 | 0.935 | 0.944 | 1.000 | 0.935 |
| $Pe_c$ | 0.753 | 0.694 | 0.620 | 0.642 | 0.978 | 0.654 | 0.936 | 0.915 | 1.000 | 0.936 |
| $Po_c - Pe_c$ | 0.145 | 0.130 | 0.075 | 0.117 | -0.006 | 0.059 | 0.000 | 0.029 | 0.000 | 0.000 |
| $1 - Pe_c$ | 0.247 | 0.306 | 0.380 | 0.358 | 0.022 | 0.346 | 0.064 | 0.085 | 0.000 | 0.064 |
| $\kappa_c$ | 0.588 | 0.426 | 0.197 | 0.327 | -0.264 | 0.170 | -0.006 | 0.346 | NaN | -0.006 |

**Supplementary Information** The online version contains supplementary material available at https://doi.org/10.3758/s13428-025-02746-8.

**Acknowledgements** The first author warmly thanks Professor Jan De Neve for his guidance during his master's thesis, conducted as part of the MSc in Statistical Data Analysis at Ghent University. This paper is based on that thesis, and his sharp statistical insight, inspiring teaching, and gentle mentorship were invaluable throughout.

**Author Contributions** • Filip Moons: Conceptualization, Formal analysis, Investigation, Project administration, Writing–original draft.
• Ellen Vandervieren: Writing–review & editing, Supervision, Funding acquisition.

**Funding** The Research Foundation of Flanders (Belgium) funded this work with a doctoral fellowship, strategic basic research 1S95920N.

**Data Availability** All materials and analysis code are available as Supplementary material. We provide an Excel spreadsheet containing all examples discussed in the paper, as well as an R script with the required data files. A detailed proof of Theorem 1 is also included. For the most recent version of the materials, additional code (e.g., for additional statistical programs), and updates, please refer to the OSF project: https://osf.io/q5nft/.

**Code Availability** All materials and analysis code are available as Supplementary material. We provide an Excel spreadsheet containing all examples discussed in the paper, as well as an R script with the required data files. A detailed proof of Theorem 1 is also included. For the most recent version of the materials, additional code (e.g., for additional statistical programs), and updates, please refer to the OSF project: https://osf.io/q5nft/.

## Declarations

**Ethical Approval** Not applicable.

**Conflicts of Interest** We have no conflicts of interest to disclose.

**Consent to Participate** Not applicable.

**Consent for Publication** Not applicable.

**Open practice statement** All materials and analysis code are available as Supplementary material. We provide an Excel spreadsheet containing all examples discussed in the paper, as well as an R script with the required data files. A detailed proof of Theorem 1 is also included. For the most recent version of the materials, additional code (e.g., for additional statistical programs), and updates, please refer to the OSF project: https://osf.io/q5nft/. None of the reported studies were preregistered. An earlier version of this manuscript has been published as an arXiv preprint prior to formal peer review and publication: https://doi.org/10.48550/arXiv.2303.12502.